%% file: main.tex
\documentclass[letterpaper]{article} % DO NOT CHANGE THIS
\usepackage[preprint]{aaai2027}  % DO NOT CHANGE THIS
\usepackage{amsmath}
\usepackage{amssymb}
\usepackage{comment}
\usepackage{xr}

\usepackage{color}

\usepackage[hyphens]{url}  % DO NOT CHANGE THIS
\usepackage{graphicx} % DO NOT CHANGE THIS
\usepackage{natbib}  % DO NOT CHANGE THIS AND DO NOT ADD ANY OPTIONS TO IT
\usepackage{caption} % DO NOT CHANGE THIS AND DO NOT ADD ANY OPTIONS TO IT
\usepackage{multirow}
\usepackage{algorithm}
\usepackage{algorithmic}

\usepackage{newfloat}
\usepackage{listings}
\DeclareCaptionStyle{ruled}{labelfont=normalfont,labelsep=colon,strut=off} % DO NOT CHANGE THIS
\floatstyle{ruled}
\newfloat{listing}{tb}{lst}{}
\floatname{listing}{Listing}

\usepackage{booktabs}

\title{TAGR: Task-Adaptive Geo-Semantic Retrieval across Heterogeneous Urban Delivery Tasks}
\author{
  Zhi Liu\textsuperscript{\rm 1},
  Yurun Cai\textsuperscript{\rm 1},
  Xiao Han\textsuperscript{\rm 1}\corresponding,
  Zhi Zheng\textsuperscript{\rm 2},
  Yuran Li\textsuperscript{\rm 1},
  Guojiang Shen\textsuperscript{\rm 1},
  Xiangjie Kong\textsuperscript{\rm 1}
}
\affiliations{
  \textsuperscript{\rm 1}Zhejiang University of Technology\\
  \textsuperscript{\rm 2}University of Science and Technology of China
}

\begin{document}

\maketitle

\begin{abstract}
As urban residents demand higher-quality instant delivery services, demand forecasting, estimated time of arrival 
prediction, and next-stop prediction have become critical tasks in urban delivery systems.
However, the three tasks rely on historical evidence with substantially different structures and granularities.
Meanwhile, their prediction targets differ in output form, making it difficult to integrate retrieved evidence through a single fusion mechanism without causing mismatch or negative transfer.
These issues can increase prediction errors and subsequently lead to imbalanced resource allocation, where some regions experience unmet demand.
In contrast, others suffer from underutilization, ultimately degrading overall efficiency and service quality.
To address these challenges, we propose TAGR, a Task-Adaptive Geo-Semantic Retrieval framework across heterogeneous urban delivery tasks.
TAGR constructs an LLM-based geo-semantic shared prior and task-specific retrieval banks to share regional knowledge while preserving heterogeneous task-specific evidence.
The LLM-derived regional representations encode geographic context and semantic relations among regions, providing transferable knowledge for all three tasks without requiring online LLM inference.
It further introduces a task-adaptive retrieval augmentation module
to accommodate 
heterogeneous output structures, thereby 
mitigating
information mismatch and negative transfer caused by a uniform retrieval-fusion strategy.

Experiments on three city 
datasets demonstrate that TAGR consistently outperforms 
state-of-the-art
baselines
across multiple tasks.

\end{abstract}

\input{sections/01_Introduction}

\input{sections/02_related_work}
\input{sections/03_Problem_Formulation}
\input{sections/04_Methodology}

\input{sections/05_Experiments}

\input{sections/06_Conclusion}

\bibliography{references}
\end{document}

%% file: sections/01_Introduction.tex
\section{Introduction}
In urban instant delivery scenarios, demand forecasting, estimated time of arrival (ETA) prediction, and next-stop prediction constitute three core spatio-temporal tasks that underpin the operational optimization of delivery systems~\cite{hess2021realtime,wang2018deeptte,liu2016strnn,wen2024rtpsurvey}. They correspond to different stages of the delivery workflow:
regional demand 
affects workload allocation, delivery conditions influence order completion times, and the 
current delivery state guides subsequent routing decisions.
Although these tasks share geographic and operational context, the historical evidence they rely on differs substantially in structure and granularity.
In addition, their prediction targets also vary in output form.
This heterogeneity makes it challenging to exploit their shared geographic and operational context without compromising task-specific modeling.

Existing approaches generally follow two paradigms: independent task-specific modeling and joint shared-representation modeling.
The former trains a dedicated model for each task and therefore captures task-specific patterns independently, whereas the latter employs a shared encoder or unified representation to extract common knowledge across tasks. However, when the
tasks differ substantially in data granularity and output form, excessive representation sharing may obscure task-specific information and introduce cross-task interference~\cite{misra2016crossstitch,liu2019mtan}. Therefore, the central distinction between the two paradigms lies in
how they balance task-specific modeling against cross-task knowledge
sharing.

\begin{figure}[t]
    \centering
    \includegraphics[width=0.90\columnwidth]
    {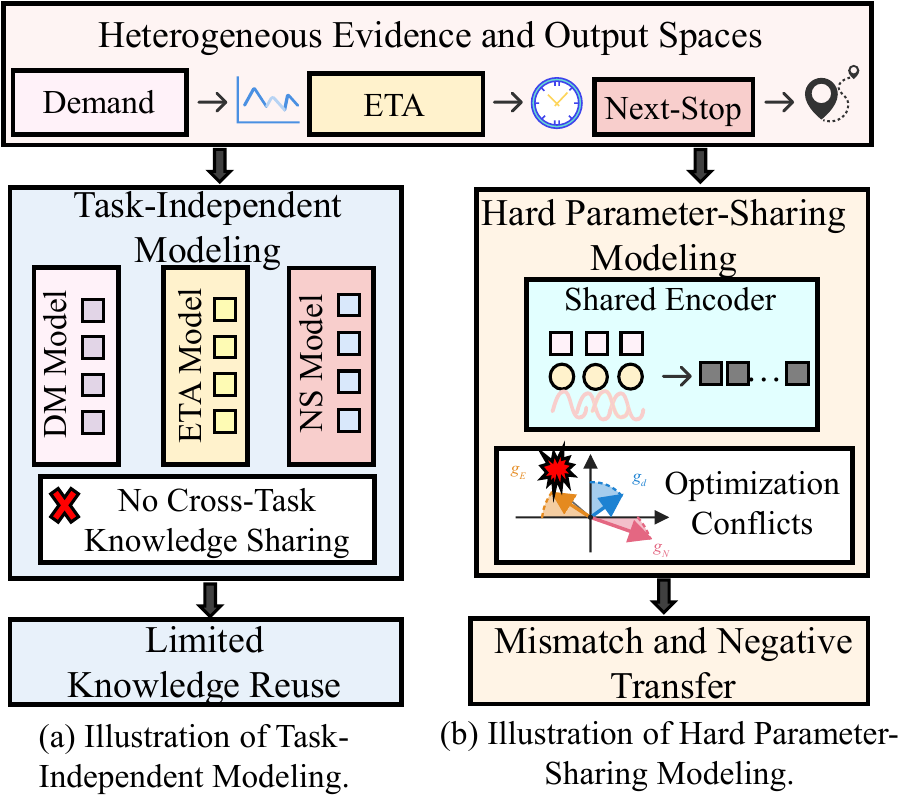}
    \vspace{-1.5mm}
    \caption{Comparison of independent single-task modeling
    and hard parameter-sharing multi-task modeling.}
    \label{fig:intro}
\end{figure}

Task-independent modeling 
employs a separate model architecture 
for each task according to its data structure, feature composition, and prediction target. This preserves the individual characteristics of each task.
As illustrated in Figure~\ref{fig:intro}(a), demand forecasting models temporal demand sequences, ETA prediction relies on order attributes and geographic context, and next-stop prediction considers route prefixes and dynamic candidate sets.
Such task-specific architectures can preserve the distinctive characteristics of each task and avoid forcing heterogeneous information into a unified representation.
However, the isolated models provide no effective channel for cross-task knowledge exchange.
Consequently, shared regional semantics, delivery patterns, and complementary historical evidence cannot be effectively transferred or reused, limiting the overall representation capacity of multi-task delivery prediction{~\cite{caruana1997multitask}.}

Joint multi-task modeling instead introduces a shared encoder or representation space to facilitate information exchange and improve parameter efficiency.
However, as illustrated in Figure~\ref{fig:intro}(b), the three tasks rely on historical evidence with substantially different structures and granularities.
Their outputs also take distinct forms, including multi-step demand sequences, scalar arrival times, and rankings over dynamic candidate sets.
Hard parameter-sharing methods compress these heterogeneous signals into a largely unified representation, despite their different information requirements and optimization objectives. 
Consequently, the learned representation may favor certain tasks but deviate from others, causing optimization conflicts, information mismatch, and negative transfer that undermine task-specific prediction~\cite{misra2016crossstitch,liu2019mtan,sener2018multitask,yu2020pcgrad}.

To address these challenges, we propose TAGR,  a unified framework
for heterogeneous urban delivery tasks.
TAGR first constructs a geo-semantic shared prior to capture reusable regional knowledge across tasks.
Specifically, a shared frozen LLM encoder maps task-oriented regional descriptions into a common semantic space, thereby establishing a unified geographic knowledge interface.
However, shared semantics alone cannot preserve the distinct historical evidence required by different tasks.
TAGR therefore introduces task-specific adapters and retrieval banks.
The adapters retain information relevant to each prediction objective, and the retrieval banks store historical evidence in the structures and granularities required by demand forecasting, ETA prediction, and next-stop prediction.
Together, these components enable cross-task knowledge reuse without obscuring task-specific patterns, providing the basis for subsequent task-adaptive retrieval and output-aligned fusion.

\begin{itemize}
\item We propose TAGR, 
a retrieval-augmented framework that jointly models ETA and next-stop prediction while extending a unified geo-semantic prior and task-adaptive retrieval mechanism to demand forecasting, thereby supporting all three core tasks in urban instant delivery.

\item Two synergistic modules are designed:
an LLM-based geo-semantic prior with task-specific retrieval banks to share regional knowledge and preserve heterogeneous evidence, and a task-adaptive
retrieval augmentation module to match each task’s granularity and output form, reducing noise and negative transfer.

\item  We conduct comprehensive multi-task delivery prediction experiments on three real-world 
city datasets spanning the scale range of the benchmark.
The experimental results demonstrate the effectiveness of TAGR across multiple cities and tasks, and ablation studies further verify the contribution of each module.
\end{itemize}

%% file: sections/02_related_work.tex
\section{Related Work}

\textbf{Urban Delivery Prediction.}
Demand forecasting, ETA prediction, and route/next-stop prediction constitute the core spatio-temporal tasks in last-mile delivery~\cite{wen2024rtpsurvey}. General spatio-temporal graph networks---including DCRNN, Graph WaveNet, MTGNN, STGCN, and AGCRN---provide strong region-level forecasting backbones~\cite{li2018dcrnn,wu2019graphwavenet,wu2020mtgnn,yu2018stgcn,bai2020agcrn}, while classical heuristics such as nearest-neighbor regression, mobility Markov chains, and constraint-programming routing solvers remain competitive references for ETA and next-stop prediction~\cite{cover1967nn,gambs2012markov,ortools}. Delivery-specific models such as DeepRoute and Graph2Route instead predict the courier's visiting order directly~\cite{wen2021deeproute,wen2022graph2route}. Beyond task-specific spatio-temporal predictors, recent studies have
explored general-purpose time-series and urban spatio-temporal
foundation models, including MOMENT and Timer-XL
~\cite{goswami2024moment,liu2025timer}.
Recent surveys have further systematized the emerging landscape of
spatio-temporal foundation models
~\cite{liang2025foundation}. 
However, existing studies typically address these tasks in isolation and do not jointly cover demand, ETA, and next-stop prediction.

\textbf{Retrieval-Augmented Spatio-Temporal Learning.}
Retrieval augmentation combines parametric models with non-parametric external knowledge~\cite{lewis2020rag}, and has recently been applied to spatio-temporal prediction, e.g., RAST for traffic forecasting~\cite{ruan2026rast} and retrieval-augmented time-series forecasters~\cite{liu2024ratd,jiang2024ragraph}. These two lines of work remain largely separate: delivery-oriented multi-task models such as M2G4RTP~\cite{cai2023m2g4rtp} jointly predict route order and arrival time but do not retrieve external evidence, whereas retrieval-augmented spatio-temporal models target homogeneous continuous-valued targets and do not handle order-level regression together with ranking over a dynamic candidate set. TAGR is, to our knowledge, the first framework to build task-adaptive geo-semantic retrieval for heterogeneous delivery tasks whose input structures, prediction granularities, and output spaces all differ, injecting evidence at output-compatible fusion stages.

%% file: sections/03_Problem_Formulation.tex
\section{Problem Formulation}

A city is divided into $N$ service regions
$\mathcal{R}=\{r_1,\ldots,r_N\}$, represented by a region
graph $\mathcal{G}=(\mathcal{R},\mathcal{E},\mathbf{A})$,
where $\mathbf{A}\in\mathbb{R}^{N\times N}$ describes the
spatial relationships among regions. Let
$\mathbf{D}_t=[d_{1,t},\ldots,d_{N,t}]\in\mathbb{R}^{N}$ denote
the city-wide delivery demand at time step $t$, where $d_{i,t}$
is the demand of region $r_i$.

We consider three delivery prediction tasks indexed by
$\mathcal{K}=\{\mathrm{dm},\mathrm{eta},\mathrm{ns}\}$, whose
inputs and targets are defined as
\begin{equation}
\left(\mathbf{X}^{(k)},\mathbf{Y}^{(k)}\right)=
\begin{cases}
\left(
\mathbf{D}_{t-T_h+1:t},
\mathbf{D}_{t+1:t+T_f}
\right), & k=\mathrm{dm},\\
\left(
\{\mathbf{x},a,r\},
\eta
\right), & k=\mathrm{eta},\\
\left(
\{S_{1:\ell},\mathbf{x}_{\mathrm{cur}},\mathcal{C}\},
s^{*}
\right), & k=\mathrm{ns},
\end{cases}
\end{equation}
where $\mathbf{x}$ and $a$ denote order attributes and AOI
type, respectively, $S_{1:\ell}=(s_1,\ldots,s_\ell)$ is the
visited route prefix, $\mathbf{x}_{\mathrm{cur}}$ is the current
courier state, $\mathcal{C}$ is the candidate set, and
$s^{*}=s_{\ell+1}$ is the actual next stop.

Each task is associated with task-oriented regional
geo-semantic representations $\mathbf{E}^{(k)}$ and a
retrieval bank $\mathcal{B}^{(k)}$ constructed from the
training split. Given an online query built by a task-specific
query constructor $Q_k$ from the current input $\mathbf{X}^{(k)}$
and its regional representations $\mathbf{E}^{(k)}$, the top-$K$
retrieved evidence $\mathcal{R}^{(k)}$ is obtained from
$\mathcal{B}^{(k)}$ under a task-specific similarity measure
$\delta_k$ (L2 distance or inner-product similarity over the
retrieval keys); the construction of $Q_k$ and $\delta_k$ is
detailed in the Methodology, Eq.~\eqref{eq:online_retrieval}.

TAGR learns three task-specific predictors:
\begin{equation}
\widehat{\mathbf{Y}}^{(k)}
=
f_k\left(
\mathbf{X}^{(k)},
\mathcal{G},
\mathbf{E}^{(k)},
\mathcal{R}^{(k)}
\right),
\quad k\in\mathcal{K}.
\end{equation}

For next-stop prediction, $f_{\mathrm{ns}}$ assigns a score
to each candidate in $\mathcal{C}$ and selects the
highest-scoring stop, following the standard formulation of
next-location prediction as ranking over a candidate
set ~\cite{feng2018deepmove,yang2020flashback}.

The order- and route-level tasks (ETA and next-stop) operate on a
shared per-region index space and are jointly optimized over shared
parameters $\Theta_{\mathrm{sh}}$ and task-specific parameters
$\Theta_k$, whereas demand forecasting,
being a region-level sequence task, shares the same geo-semantic
prior and retrieval mechanism but is optimized separately.

%% file: sections/04_Methodology.tex
\section{Methodology}

\subsection{Overview}
\begin{figure*}[t]
    \centering
    \includegraphics[width=0.85\textwidth]
    {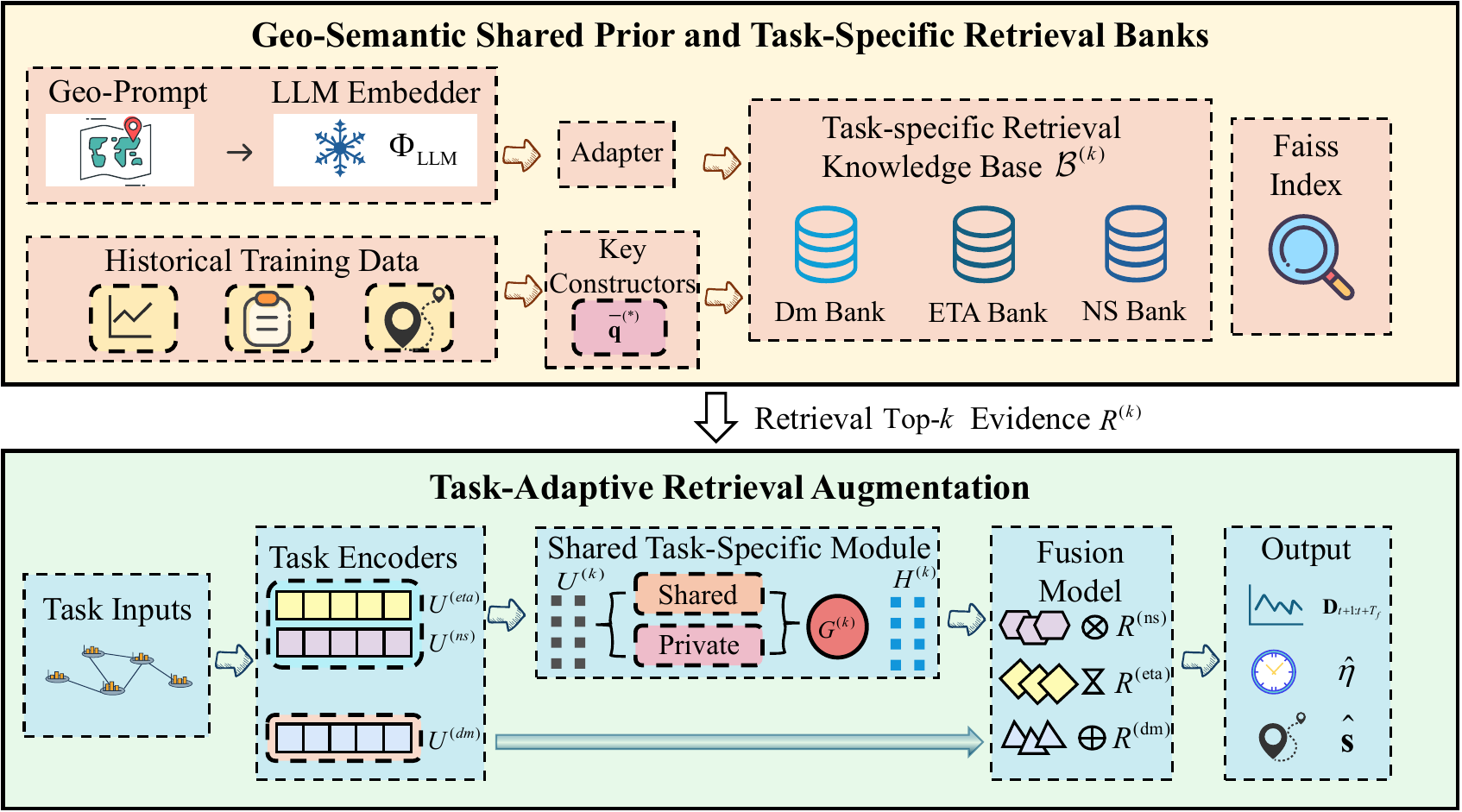}
    \caption{Overall architecture of TAGR. The offline module
    constructs geo-semantic priors and task-specific retrieval banks,
    while the online module performs task-adaptive retrieval augmentation
    for demand, ETA, and next-stop prediction.}
    \label{fig:TAGR_framework}
\end{figure*}

Figure~\ref{fig:TAGR_framework} presents the overall architecture of TAGR, which comprises two principal modules. The first module maps task-oriented regional descriptions into geo-semantic representations and constructs task-specific retrieval banks from historical demand windows, order contexts, and route prefixes, enabling cross-task sharing of regional features and delivery knowledge. The second module retrieves task-relevant historical evidence and incorporates it into the representation, prediction, and candidate spaces for demand, ETA, and next-stop prediction, respectively.

\subsection{Geo-Semantic Prior and Retrieval Banks}
Urban delivery prediction involves heterogeneous spatio-temporal data: regional demand is a multi-step temporal sequence, ETA prediction relies on order-level attributes and geographic context, and next-stop prediction depends on route prefixes, courier states, and dynamic candidate sets. A single shared representation is therefore unsuitable, since it either obscures task-specific patterns or blocks cross-task transfer. TAGR therefore constructs reusable geographic and historical knowledge through a geo-semantic shared prior that provides a common regional knowledge interface, together with task-specific retrieval banks that preserve the distinct evidence forms of demand sequences, order records, and route prefixes. Large Language Model encoding, retrieval-projection training, and index construction are all performed offline.

Specifically, each region $r_i$ is described by structured geographic attributes, and for each task $k\in\mathcal{K}$ a deterministic prompt constructor $\mathcal{T}_k$ converts the corresponding geographic information $\mathbf{g}_i^{(k)}$ into a task-oriented textual description. As in recent LLM-based urban prediction~\cite{jin2024timellm,li2024urbangpt,yuan2024unist}, a shared frozen language encoder maps the description into a common geo-semantic space, followed by a task-specific adapter:
\begin{equation}
\mathbf{z}_i^{(k)}
= \operatorname{Adapter}_k\!\left(\Phi_{\mathrm{LLM}}\!\left(\mathcal{T}_k\!\left(\mathbf{g}_i^{(k)}\right)\right)\right),
\label{eq:semantic_encoding}
\end{equation}
where $p_i^{(k)}=\mathcal{T}_k(\mathbf{g}_i^{(k)})$ is the task-oriented prompt, $\mathbf{e}_i^{(k)}=\Phi_{\mathrm{LLM}}(p_i^{(k)})$ its frozen encoding, and $\mathbf{E}^{(k)}=[\mathbf{z}_1^{(k)};\ldots;\mathbf{z}_N^{(k)}]$ stacks the per-task regional representations, with each $\operatorname{Adapter}_k$ emphasizing regional functions for demand, road accessibility for ETA, and destination characteristics for next-stop. For demand forecasting, these representations further construct a semantic support matrix $\mathbf{A}^{(\mathrm{sem})}$.

TAGR complements this shared prior with a separate key--value retrieval bank per task, built from the training split, to preserve task-specific historical evidence:
\begin{equation}
\bar{\mathbf{q}}_{i,t}^{(\mathrm{dm})}
= \psi_{\mathrm{dm}}(d_{i,t-T_h+1:t}),
\label{eq:demand_bank}
\end{equation}
where $\psi_{\mathrm{dm}}(\cdot)$ extracts temporal statistics and trend descriptors, and the paired value is the future-demand segment $\mathbf{v}_{i,t}^{(\mathrm{dm})}=d_{i,t+1:t+T_f}$. For ETA prediction, the key augments the normalized order context with an AOI-aware encoding of its destination:
\begin{equation}
\bar{\mathbf{q}}_j^{(\mathrm{eta})}
= [\widetilde{\mathbf{x}}_j;\,\tau_{\mathrm{AOI}}(a_j)],
\label{eq:eta_bank}
\end{equation}
where $\widetilde{\mathbf{x}}_j$ denotes the normalized order and delivery attributes, $\tau_{\mathrm{AOI}}(\cdot)$ a train-only AOI target encoding, and $\mathbf{v}_j^{(\mathrm{eta})}=\eta_j$ the observed completion time. For next-stop prediction, the key aggregates the visited route prefix while the value is the actual subsequent stop:
\begin{equation}
\bar{\mathbf{q}}_j^{(\mathrm{ns})}
= \frac{1}{\ell_j}\sum_{u=1}^{\ell_j}\mathbf{z}_{s_u}^{(\mathrm{ns})},
\label{eq:nextstop_bank}
\end{equation}
where $\ell_j$ is the length of the observed prefix $S_{1:\ell_j}=(s_1,\ldots,s_{\ell_j})$, $\mathbf{z}_{s_u}^{(\mathrm{ns})}$ is the next-stop-oriented representation of the $u$-th visited stop, and $\mathbf{v}_j^{(\mathrm{ns})}=\mathbf{z}_{s_{\ell_j+1}}^{(\mathrm{ns})}$ is the actual next stop $s^{*}=s_{\ell_j+1}$; this key is order-invariant, unlike the visiting order modeled by the next-stop encoder.

All three tasks retrieve directly on these base keys, the AOI-augmented order key for ETA, the demand-window descriptor $\bar{\mathbf{q}}^{(\mathrm{dm})}$, and the geo-semantic prefix aggregation $\bar{\mathbf{q}}^{(\mathrm{ns})}$, without any learned query projection. The resulting key--value pairs form the task-specific retrieval banks $\mathcal{B}^{(k)}$, whose FAISS indices are built offline for efficient L2-distance or inner-product search.

\subsection{Task-Adaptive Retrieval and Output Fusion}
Because regional demand forecasting targets region-level multi-step sequences, ETA prediction produces an order-level scalar, and next-stop prediction ranks a dynamic set of candidate locations, a single encoding and fusion operator cannot accommodate their divergent output structures. TAGR therefore introduces a task-adaptive retrieval and output-aligned fusion module that encodes task inputs through task-specific encoders and a gated shared--private representation module, retrieves evidence from the corresponding historical bank, and incorporates it into the representation, prediction, or candidate space.

For each task $k\in\mathcal{K}$, TAGR employs a task-specific encoder $\operatorname{Enc}_k(\cdot)$ to encode the structured input $\mathbf{X}^{(k)}$, followed by a projection layer that maps the heterogeneous task representations into a common hidden dimension:
\begin{equation}
\mathbf{U}^{(k)}
= \operatorname{Proj}_k\!\left(
\operatorname{Enc}_k\!\left(\mathbf{X}^{(k)},\mathcal{G},\mathbf{E}^{(k)}\right)
\right),
\quad k\in\mathcal{K}.
\label{eq:task_encoding}
\end{equation}

Because independent region representations limit cross-task transfer whereas fully shared ones may cause negative transfer, for ETA and next-stop, which share the same per-region index space, TAGR decomposes the regional representation into a task-agnostic shared component and task-specific private components and fuses them with a task-adaptive gate:
\begin{equation}
\mathbf{H}^{(k)} = \mathbf{G}^{(k)}\odot\mathbf{S} + \left(1-\mathbf{G}^{(k)}\right)\odot\mathbf{P}^{(k)},
\quad k\in\{\mathrm{eta},\mathrm{ns}\},
\label{eq:shared_private_fusion}
\end{equation}
where $\mathbf{U}^{\mathrm{sh}}\in\mathbb{R}^{N\times d}$ is a task-agnostic per-region feature shared by ETA and next-stop, $\mathbf{S}=F_{\mathrm{sh}}(\mathbf{U}^{\mathrm{sh}})$ and $\mathbf{P}^{(k)}=F_{\mathrm{pr}}^{(k)}(\mathbf{U}^{(k)})$ transform the shared and task-specific features, and the gate $\mathbf{G}^{(k)}=\sigma(\operatorname{MLP}_{g}^{(k)}(\mathbf{S}\Vert\mathbf{P}^{(k)}))$ is computed per region and feature dimension, yielding the prediction backbone $\mathbf{H}^{(k)}$.
Demand bypasses this module, instead sharing cross-region knowledge through the semantic support matrix $\mathbf{A}^{(\mathrm{sem})}$, with $\mathbf{H}^{(\mathrm{dm})}$ produced directly by the demand encoder. Each task then constructs its online query with the same key map used to build its bank, instantiating the query constructor $Q_k$ of the problem formulation:
\begin{equation}
\mathcal{R}^{(k)}
= \operatorname{TopK}(\mathbf{q}^{(k)},\mathcal{B}^{(k)};\delta_k),\quad k\in\mathcal{K},
\label{eq:online_retrieval}
\end{equation}
where $\mathbf{q}^{(k)}=Q_k^{\mathrm{base}}(\mathbf{X}^{(k)},\mathbf{E}^{(k)})$ for $k\in\mathcal{K}$, with the ETA query reusing its AOI-augmented order key while demand and next-stop use their base keys.

Demand queries depend only on the current historical demand window, whereas ETA and next-stop queries also incorporate regional semantic representations. Validation and test samples are retrieved only from banks built on the training split, preventing information leakage.

We therefore map the retrieved future-demand segments into the hidden representation $\mathbf{R}^{(\mathrm{dm})}=[\phi_{\mathrm{dm}}(\mathbf{v}_1^{(\mathrm{dm})});\ldots;\phi_{\mathrm{dm}}(\mathbf{v}_K^{(\mathrm{dm})})]$, and use the current demand representation as the query while the retrieved segments serve as keys and values in a multi-head attention mechanism:
\begin{equation}
\widetilde{\mathbf{H}}^{(\mathrm{dm})}
= \mathbf{H}^{(\mathrm{dm})} + \sigma\!\left(\gamma_{\mathrm{dm}}\right)\mathbf{M}^{(\mathrm{dm})},
\label{eq:demand_augmentation}
\end{equation}
where $\mathbf{M}^{(\mathrm{dm})}=\operatorname{MHA}(\mathbf{H}^{(\mathrm{dm})},\mathbf{R}^{(\mathrm{dm})},\mathbf{R}^{(\mathrm{dm})})$ is multi-head attention~\cite{vaswani2017attention} from the current demand representation to the retrieved segments, $\gamma_{\mathrm{dm}}$ is a single global learnable parameter scaling the contribution of historical demand evidence, and the demand output is $\widehat{\mathbf{Y}}^{(\mathrm{dm})}=\operatorname{Head}_{\mathrm{dm}}(\widetilde{\mathbf{H}}^{(\mathrm{dm})})$. ETA instead aggregates retrieved completion times directly in the prediction space, as a locally distance-weighted average: \begin{equation}
\eta_{\mathrm{ret}} = \sum_{j=1}^{K}\omega_j\,\mathbf{v}_j^{(\mathrm{eta})},
\label{eq:eta_retrieval}
\end{equation}
where $d_j$ is the distance of the $j$-th retrieved order from the query, $\omega_j=\exp(-d_j/\tau_{\mathrm{eta}})/\sum_{u=1}^{K}\exp(-d_u/\tau_{\mathrm{eta}})$ are the resulting softmax weights, and, with the base estimate $\widehat{\eta}_{\mathrm{base}}=\operatorname{Head}_{\mathrm{eta}}(\mathbf{H}^{(\mathrm{eta})})$, the two are combined through a reliability gate over neighbor-distance statistics $\mathbf{r}^{(\mathrm{eta})}$:
\begin{equation}
\widehat{\eta}
= \left(1-g_{\mathrm{eta}}\right)\widehat{\eta}_{\mathrm{base}}
+ g_{\mathrm{eta}}\,\eta_{\mathrm{ret}},
\label{eq:eta_augmentation}
\end{equation}
where the reliability gate $g_{\mathrm{eta}}=\sigma(\operatorname{MLP}_{g}^{(\mathrm{eta})}(\mathbf{H}^{(\mathrm{eta})}\Vert\mathbf{r}^{(\mathrm{eta})}))$ weights the retrieval estimate against the parametric one.

TAGR first aggregates the actual subsequent-stop representations from the retrieved historical routes according to their similarities, producing a historical route prior $\mathbf{h}_{\mathrm{ret}}$ whose agreement with each candidate $s_m^{e}$ yields a prior score $\pi_m$:
\begin{equation}
\mathbf{h}_{\mathrm{ret}}
= \sum_{j=1}^{K}\beta_j\mathbf{v}_j^{(\mathrm{ns})},
\label{eq:nextstop_prior}
\end{equation}
where $\beta_j=\exp(\operatorname{sim}_j)/\sum_{u=1}^{K}\exp(\operatorname{sim}_u)$ are the similarity-based aggregation weights over the retrieved routes, and the prior score of candidate $s_m^{e}$ is its cosine agreement with the route prior, $\pi_m=\cos(\mathbf{z}_{s_m^{e}}^{(\mathrm{ns})},\mathbf{h}_{\mathrm{ret}})$.

To assess whether the prior discriminates among candidates, we summarize its dispersion into a compact statistic $\mathbf{r}_{\pi}$, where large dispersion indicates a clear preference while nearly uniform scores imply limited ranking information. The reliability gate and candidate prior are then jointly incorporated into candidate scoring:
\begin{equation}
\rho_m = \operatorname{MLP}_{\mathrm{ptr}}\!\left(\mathbf{o}_m^{(\mathrm{ns})}\right) + g_{\mathrm{ns}}\gamma_{\mathrm{ns}}\pi_m,
\label{eq:nextstop_augmentation}
\end{equation}
where the reliability gate $g_{\mathrm{ns}}=\sigma(\operatorname{MLP}_{g}^{(\mathrm{ns})}(\mathbf{r}_{\pi}))$ is driven by the prior dispersion, $\mathbf{o}_m^{(\mathrm{ns})}=\mathbf{H}^{(\mathrm{ns})}\Vert\mathbf{c}_m\Vert g_{\mathrm{ns}}\mathbf{h}_{\mathrm{ret}}$ is the fused candidate representation with $\mathbf{c}_m$ the representation of candidate $s_m^{e}\in\mathcal{C}$, and $\gamma_{\mathrm{ns}}$ controls the contribution of the historical prior score.

TAGR trains all modules jointly in a single stage after building the retrieval banks offline. The two tasks use Huber loss for ETA regression and cross-entropy over valid candidates for next-stop ranking; given the fused ETA prediction $\hat{\eta}$ in Eq. (12) and the observed completion time $\eta$, the ETA loss is
\begin{equation}
\mathcal{L}_{\mathrm{eta}} =
\begin{cases}
\frac{1}{2}(\hat{\eta}-\eta)^2, & |\hat{\eta}-\eta| < 5,\\
5\left(|\hat{\eta}-\eta|-\frac{1}{2}\cdot 5\right), & \text{otherwise},
\end{cases}
\label{eq:eta_loss}
\end{equation}
where the threshold is set to 5 minutes in implementation. For next-stop prediction, Eq. (14) produces the candidate score $\rho_m$ for each candidate $e_m \in C$. After visited, invalid, and padded candidates are masked, the next-stop loss is
\begin{equation}
\mathcal{L}_{\mathrm{ns}}
= -\log
\frac{\exp(\rho_m)\big|_{e_m=s^*}}
{\sum_{e_m \in C} \exp(\rho_m)} .
\label{eq:ns_loss}
\end{equation}

The two losses are averaged over their mini-batches and combined with fixed equal weights~\cite{kendall2018multitask}:
\begin{equation}
\mathcal{L}_{\mathrm{joint}}
= \mathcal{L}_{\mathrm{eta}} + \mathcal{L}_{\mathrm{ns}}.
\label{eq:joint_loss}
\end{equation}

Each training step draws one mini-batch from each dataset and combines the two losses in a single backward pass, so shared parameters receive gradients from both tasks while task-specific modules update only on their own task. Demand forecasting is trained separately under a masked MAE loss $\mathcal{L}_{\mathrm{dm}}$.

%% file: sections/05_Experiments.tex
\section{Experiments}

\paragraph{Datasets.}
We conduct experiments on the pickup subset of the LaDe dataset (LaDe-P)~\cite{wu2024lade}, an industry-scale last-mile logistics dataset collected from real-world operations. We evaluate on three city subsets that span the scale range of the dataset: HZ (Hangzhou), CQ (Chongqing), and JL (Jilin).

\paragraph{Baselines.}
We compare TAGR against baselines from three groups. Across all
three tasks, we adapt three deep spatio-temporal models, including
 DCRNN~\cite{li2018dcrnn}, GWNet~\cite{wu2019graphwavenet},
MTGNN~\cite{wu2020mtgnn} and the LLM-based
IMPEL~\cite{nie2025impel}, each using its encoder as a
shared region backbone with task-specific heads.
We further add output-form-matched baselines: a historical-average
heuristic (HA), STGCN~\cite{yu2018stgcn},
AGCRN~\cite{bai2020agcrn}, and retrieval-augmented
RAST~\cite{ruan2026rast} for demand; a mean-speed heuristic(SPEED), KNN regression,and M2G4RTP~\cite{cai2023m2g4rtp} for
ETA; and, for next-stop, a popularity heuristic (MostPop), a
mobility Markov chain~\cite{gambs2012markov}, a
nearest-neighbor Greedy rule, an OR-Tools~\cite{ortools} routing
solver, and three delivery-specific route models, including
M2G4RTP~\cite{cai2023m2g4rtp},
DeepRoute~\cite{wen2021deeproute}, and
Graph2Route~\cite{wen2022graph2route}.

\paragraph{Implementation Details.}
All results are averaged over five random seeds, and baselines use
their official settings tuned on the same validation split. Regional
profiles are encoded offline by the frozen
Qwen3-Embedding-8B model~\cite{zhang2025qwen3embedding} into
$4096$-dimensional representations, and we retrieve the top-$5$
neighbors with route prefixes of length $5$. We report
MAE and RMSE for demand and ETA, and HR@$k$ and MRR@$k$ for
next-stop.

\subsection{Performance Comparison}

\begin{table*}[t]
\centering
\small
\setlength{\tabcolsep}{3.5pt}
\renewcommand{\arraystretch}{1.15}
\begin{tabular}{@{}l cc cc ccc@{}}
\toprule
& \multicolumn{2}{c}{\textbf{Demand}} & \multicolumn{2}{c}{\textbf{ETA}} & \multicolumn{3}{c}{\textbf{Next-stop}} \\
\cmidrule(lr){2-3}\cmidrule(lr){4-5}\cmidrule(lr){6-8}
\textbf{Model}
& MAE$\downarrow$ & RMSE$\downarrow$
& MAE$\downarrow$ & RMSE$\downarrow$
& HR@5$\uparrow$ & HR@20$\uparrow$ & MRR@10$\uparrow$ \\
\midrule
GWNet   & 6.19/3.25/4.29 & 12.66/6.89/8.30 & 25.92/24.96/26.70 & 31.82/30.73/32.88 & 0.29/0.45/0.40 & 0.71/0.74/0.72 & 0.18/0.31/0.25 \\
MTGNN   & 6.06/3.13/1.43 & 12.14/6.20/2.74 & 25.84/24.92/26.73 & 31.64/30.62/32.88 & 0.32/0.46/0.57 & 0.67/0.76/0.78 & 0.21/0.32/0.39 \\
DCRNN   & 7.39/3.60/1.76 & 13.58/6.89/3.28 & 25.90/24.94/26.70 & 31.76/30.65/32.89 & 0.23/0.28/0.43 & 0.70/0.71/0.73 & 0.15/0.18/0.27 \\
HA      & 6.41/3.29/1.39 & 13.05/6.68/2.70 & -/-/- & -/-/- & -/-/- & -/-/- & -/-/- \\
STGCN   & 6.02/3.11/1.77 & 12.49/6.68/3.01 & -/-/- & -/-/- & -/-/- & -/-/- & -/-/- \\
AGCRN   & 4.18/2.56/1.43 & 8.19/5.23/3.41 & -/-/- & -/-/- & -/-/- & -/-/- & -/-/- \\
RAST    & 5.08/2.53/1.64 & 9.38/4.95/3.04 & -/-/- & -/-/- & -/-/- & -/-/- & -/-/- \\
SPEED   & -/-/- & -/-/- & 42.26/37.21/34.94 & 52.56/47.18/44.93 & -/-/- & -/-/- & -/-/- \\
KNN     & -/-/- & -/-/- & 27.48/26.63/27.42 & 32.84/31.82/32.53 & -/-/- & -/-/- & -/-/- \\
MostPop & -/-/- & -/-/- & -/-/- & -/-/- & 0.16/0.38/0.20 & 0.55/0.69/0.75 & 0.10/0.24/0.11 \\
Markov  & -/-/- & -/-/- & -/-/- & -/-/- & 0.66/0.52/0.30 & 0.95/0.95/0.87 & 0.46/0.35/0.20 \\
Greedy  & -/-/- & -/-/- & -/-/- & -/-/- & 0.09/0.18/0.30 & 0.37/0.52/0.59 & -/-/- \\
OR-Tools & -/-/- & -/-/- & -/-/- & -/-/- & 0.11/0.14/0.17 & 0.41/0.39/0.41 & -/-/- \\
M2G4RTP & -/-/- & -/-/- & \textbf{23.35}/25.91/\textbf{24.46} & 33.63/38.75/36.50 & 0.15/0.13/0.13 & 0.56/0.52/0.52 & -/-/- \\
DeepRoute & -/-/- & -/-/- & -/-/- & -/-/- & 0.45/0.54/0.64 & 0.77/0.86/0.89 & 0.33/0.37/0.46 \\
Graph2Route & -/-/- & -/-/- & -/-/- & -/-/- & 0.48/0.61/0.64 & 0.80/0.86/0.88 & 0.36/0.44/0.46 \\
IMPEL   & 4.52/2.47/1.39 & 9.90/4.92/2.51 & 25.87/24.97/26.39 & 31.99/30.85/32.59 & 0.40/0.50/0.76 & 0.75/0.79/0.83 & 0.28/0.35/0.27 \\
\textbf{TAGR}
& \textbf{3.98/2.40/1.33} & \textbf{7.10/4.65/2.46} & 25.73/\textbf{24.85}/26.04 & \textbf{31.53/30.60/32.17}
& \textbf{0.74/0.77/0.76} & \textbf{0.97/0.98/0.98} & \textbf{0.57/0.56/0.57} \\
\midrule
\shortstack{\textbf{Improve}\\[-0.4ex]\textbf{(\%)}}
& \shortstack[c]{4.78/2.83/4.32}
& \shortstack[c]{13.31/5.49/1.99}
& \shortstack[c]{-10.19/0.28/-6.46}
& \shortstack[c]{0.35/0.07/1.11}
& \shortstack[c]{12.16/26.68/0.40}
& \shortstack[c]{1.47/3.27/9.42}
& \shortstack[c]{23.64/25.68/23.23}  \\
\bottomrule
\end{tabular}
\caption{Performance comparison with different baselines. Values in each cell follow
the order HZ/CQ/JL. A dash (``$-$'') indicates that a task-specific
baseline is not applicable.}
\label{tab:main_results}
\end{table*}

As shown in Table~\ref{tab:main_results}, TAGR achieves the best or near-best performance across almost all tasks and metrics on all three datasets. Compared with conventional spatio-temporal models (DCRNN, GWNet, MTGNN, STGCN, AGCRN), TAGR achieves stronger demand forecasting: these models capture temporal patterns and fixed spatial dependencies purely through graph convolution or recurrent structures, and cannot explicitly represent regional functions, road accessibility, or delivery-related semantics that vary across regions, whereas TAGR complements them with an LLM-derived geo-semantic prior that supplies exactly this missing information, confirming the value of combining geographic semantics with spatio-temporal modeling for urban delivery prediction. Compared with heuristic and non-parametric baselines (HA, SPEED, KNN, MostPop, Markov, Greedy, OR-Tools), TAGR is substantially more robust across tasks: these methods rely on static statistics, local similarity, or hand-crafted rules, so a fixed rule or single similarity measure cannot adapt to complex order attributes, dynamic candidate sets, and the differing output forms of the three tasks, whereas TAGR learns task-specific representations and dynamically fuses retrieved evidence, showing that adaptive, learnable retrieval is necessary where static heuristics fall short. Compared with RAST and IMPEL, which already combine learning with retrieval or semantic representations and are thus the closest comparisons to TAGR, TAGR still outperforms both on most metrics: RAST applies a relatively uniform retrieval-augmentation strategy that cannot simultaneously adapt to demand regression, ETA regression, and next-stop ranking, and IMPEL uses semantic representations only as static input features without task-specific historical retrieval, whereas TAGR injects retrieved evidence into the representation, prediction, and candidate spaces according to each task's output form, underscoring the importance of output-aligned retrieval fusion under heterogeneous task targets.

The results further reveal two finer-grained patterns. For ETA prediction, TAGR obtains the best RMSE in all three cities, though its MAE is not always the lowest. Since RMSE is more sensitive to large errors and long-tail orders, this indicates that the Huber loss and retrieval-reliability gate mainly suppress large deviations on abnormal completion times rather than uniformly lowering average error, showing that ETA prediction benefits as much from robustness to outliers as from average accuracy. For next-stop prediction, TAGR achieves the most significant gains of the three tasks. Existing route models capture transition patterns or sequential dependencies but lack explicit retrieval of similar historical routes to form a candidate-level prior, whereas TAGR converts retrieved future stops into a route prior that directly enhances candidate scoring, confirming the particular value of historical-route retrieval for next-stop ranking.
\subsection{Ablation Study}

We conduct ablation experiments on the Hangzhou dataset to evaluate geo-semantic representations and task-adaptive retrieval. Results for all three tasks are reported in Table~\ref{tab:ablation}.

\begin{table}[t]
\centering
\small
\setlength{\tabcolsep}{3.0pt}
\renewcommand{\arraystretch}{0.95}

\begin{tabular}{@{}llcccc@{}}
\toprule
\textbf{Task}
& \textbf{Metric}
& \textbf{TAGR}
& \textbf{w/o Sem.}
& \textbf{w/o RAG}
& \textbf{Uniform} \\
\midrule

\multirow{2}{*}{Demand}
& MAE$\downarrow$
& \textbf{3.98}
& 6.39
& 4.41
& \underline{4.11} \\

& RMSE$\downarrow$
& \textbf{7.10}
& 11.06
& 7.92
& \underline{7.41} \\

\multirow{2}{*}{ETA}
& MAE$\downarrow$
& \textbf{25.73}
& 26.01
& 25.91
& \underline{25.88} \\

& RMSE$\downarrow$
& \textbf{31.53}
& 32.32
& 32.06
& \underline{31.98} \\

\multirow{3}{*}{Next-stop}
& HR@5$\uparrow$
& \textbf{0.74}
& \underline{0.64}
& 0.40
& 0.56 \\

& HR@20$\uparrow$
& \textbf{0.97}
& \underline{0.92}
& 0.71
& 0.86 \\

& MRR@10$\uparrow$
& \textbf{0.57}
& \underline{0.49}
& 0.29
& 0.43 \\

\bottomrule
\end{tabular}

\caption{Ablation results on Hangzhou.}
\label{tab:ablation}
\end{table}

\paragraph{Effect of the Geo-Semantic Prior and Retrieval Banks.}
As shown in Table~\ref{tab:ablation}, We evaluate two variant 
\textit{w/o Sem. Rep.}, which replaces LLM-derived representations with trainable region-ID embeddings, and \textit{w/o RAG},which removes all retrieval components. Removing the semantic prior hurts demand the most, while removing retrieval hurts next-stop the most; ETA is relatively insensitive to either. This is because the semantic prior supplies transferable, region-level categorical knowledge, whereas the retrieval bank supplies sample-level, continuously updated historical evidence, which next-stop prediction relies on to ground candidate scoring in precedent routes. This confirms that the two modules are not redundant: each covers a distinct historical-knowledge requirement, and removing either leaves the corresponding task without its primary information source. 

\begin{table}[t]
\centering
\small
\setlength{\tabcolsep}{2.5pt}
\renewcommand{\arraystretch}{0.94}

\begin{tabular}{@{}lccc@{}}
\toprule
\textbf{Metric}
& \textbf{Task-indep.}
& \textbf{Shared-private}
& \textbf{TAGR} \\
\midrule
Parameters
& 3.16M
& 3.21M
& 3.28M \\
ETA MAE$\downarrow$
& 25.91
& 25.88
& \textbf{25.73} \\
ETA RMSE$\downarrow$
& 31.82
& 31.78
& \textbf{31.53} \\
NS HR@5$\uparrow$
& 0.72
& 0.73
& \textbf{0.74} \\
NS HR@20$\uparrow$
& 0.87
& 0.96
& \textbf{0.97} \\
NS MRR@10$\uparrow$
& 0.56
& 0.55
& \textbf{0.57} \\
\bottomrule
\end{tabular}

\caption{Comparison of multi-task structures on Hangzhou.}
\label{tab:multitask_structures}
\end{table}

\paragraph{Effect of Task-Adaptive Retrieval Fusion.}
With both modules intact, Uniform replaces the sample-adaptive reliability gates (geta, gns) with fixed weights (e.g., a 0.5/0.5 combination of the retrieval estimate and the parametric prediction for ETA). Uniform is clearly worse than TAGR but still better than w/o RAG, showing that retrieval with poorly matched fusion still beats no retrieval at all, but falls well short of retrieval's full benefit. This is because a fixed weight cannot determine whether the evidence retrieved for a given query is close and reliable or distant and weakly related, leading to under-reliance when the evidence is trustworthy and over-reliance on noise when it is not. This effect is most evident in next-stop prediction, where candidate scoring is highly sensitive to evidence quality. This confirms that the value of retrieval augmentation depends not only on whether evidence is retrieved, but on whether its reliability can be assessed and weighted adaptively, validating the need for output-aligned, sample-adaptive fusion.

We further compare task-independent modeling, shared--private
modeling without gating, and the full gated TAGR on Hangzhou
(Table~\ref{tab:multitask_structures}). Ungated shared-private fusion outperforms task-independent modeling, showing that a shared component enables knowledge exchange between the two tasks; yet it still falls short of gated TAGR. This is because task-independent modeling fully isolates the regional knowledge accumulated by each task, while ungated sharing, although it introduces a shared component, combines the shared and private information through a fixed operator that cannot adapt when the two tasks require different sharing ratios on a given sample. This shows that the relationship between ETA and next-stop cannot be reduced to whether to share representations, but also depends on how much to share on a per-sample basis, confirming the necessity of the gated design over both ungated sharing and independent modeling.

\subsection{Sensitivity Analysis}
\label{sec}
\begin{figure}[t]
\centering
\includegraphics[width=\columnwidth]{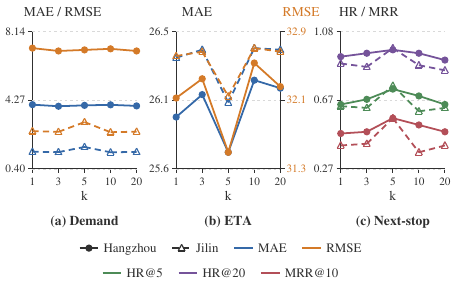}
\caption{Sensitivity of TAGR to the number of retrieved neighbors on Hangzhou and Jilin across three tasks.}
\label{fig}
\end{figure}

We vary the number of retrieved neighbors $K$ over $\{1,3,5,10,20\}$ on Hangzhou and Jilin, with all other settings fixed. The results are shown in Figure~\ref{fig}.

Increasing the retrieval size does not consistently improve performance. The benefit of retrieval comes from relevant historical evidence rather than the number of neighbors; overly large $K$ introduces weakly related samples and noise. Thus, retrieval size should be selected according to task-specific evidence structures. Demand and ETA generally prefer small or moderate $K$. Both tasks depend on locally similar demand patterns or order contexts, while excessive neighbors may include inconsistent trends or completion-time distributions. This indicates that regression tasks require precise retrieval rather than broad retrieval. Next-stop prediction performs best around $K$=5. A moderate number of similar routes provides a clear route prior, whereas too many routes make the candidate prior more dispersed and less discriminative. This shows that next-stop ranking is especially sensitive to the relevance of retrieved route evidence.

%% file: sections/06_Conclusion.tex
\section{Conclusion}
In this paper, we propose TAGR, a geo-semantic and
task-adaptive retrieval-augmented framework for urban delivery
prediction. Experiments on three real-world city datasets
demonstrate consistent improvements in demand forecasting,
ETA estimation, and next-stop prediction. In future work, we will explore adaptive
retrieval sizes, online knowledge-base updates, and evaluations
in more cities and evolving delivery environments.